\documentclass[twocolumn,showpacs,preprintnumbers,amsmath,amssymb]{revtex4}

\usepackage{graphicx}
\usepackage{dcolumn}
\usepackage{bm}

\begin{document}

\newcommand{\ra}{\rangle}
\newcommand{\la}{\langle}

\title{Environmental noise reduction for holonomic quantum gates}

\author{Daniele Parodi,$^{1,2}$ Maura Sassetti,$^{1,3}$
Paolo Solinas,$^{4}$  and Nino Zangh\`{\i}$^{1,2}$
}

\affiliation{
$^1$ Dipartimento di Fisica, Universit\`a di Genova, Genova, Italy \\
$^2$ Istituto Nazionale di Fisica Nucleare (Sezione di Genova), Genova, Italy 
$^3$ INFM-CNR Lamia \\
Via Dodecaneso 33, 16146 Genova, Italy \\
$^4$ Laboratoire de Physique Th\'eorique de la Mati\`ere Condens\'ee,
     Universit\'e Pierre et Marie Curie, Place Jussieu, 
     75252 Paris Cedex 05, France
}

\date{\today}

\begin{abstract}

We  study the  performance of holonomic quantum gates, driven by lasers, under the effect of a dissipative environment  modeled as a thermal bath of  oscillators.  
We show how to enhance the performance of the gates by suitable choice of the loop in the manifold  of  the controllable parameters of the laser. For a simplified, albeit realistic model, we find the surprising result that for a long time evolution the performance of the gate (properly estimated in terms of average fidelity) increases.  On the basis of this result, we compare holonomic   gates with the so-called Stimulated Raman adiabatic passage (STIRAP) gates.

\end{abstract}


\maketitle

\section{Introduction}

The major challenge for quantum computation is posed by the fact that generically quantum states are very delicate objects quite  difficult to control with the required accuracy---typically, by means of external driving fields, e.g., a laser. 
The interaction with  the many degrees of freedom of the environment causes  
decoherence; moreover, errors in processing the 
information may lead to a wrong output state. 

Among the approaches aiming at overcoming these difficulties are those for which the quantum gate depends very weakly on the details of the dynamics,   in particular,  the holonomic quantum computation (HQC) \cite{HQC}  and the so-called Stimulated Raman adiabatic passage (STIRAP) \cite{Kis, troiani-molinari, roszak}. In the latter,  the gate operator is obtained acting on the phase difference of the driving lasers 
during the evolution, while in the  former the same goal is achieved by exploiting the non-commutative analogue of the Berry phase collected by a quantum state
during a cyclic evolution. Concrete proposals  have been put forward, for both 
Abelian  \cite{Jones,Falci} and
non-Abelian holonomies 
\cite{HQC_proposal,HQC_proposal1,HQC_proposal2,HQC_proposal3,HQC_proposal4,paper1-2}. 
The main advantage of the HQC is the robustness 
against noise deriving from a imperfect control of the driving fields
\cite{par_noise,par_noise1,par_noise2,par_noise3,par_noise4,par_noise5,florio,fuentes}.

In a recent paper  \cite{hqc_noise} we have shown that  the disturbance of the environment on holonomic gates can be suppressed and   the performance of the gate  
optimized for  particular environments (purely superohmic thermal bath).   In the present paper we consider a different sort of optimization, which is independent of the particular nature of the environment. 

By exploiting the full geometrical structure of HQC, we show how the performance of a holonomic gate
can be enhanced by a suitable choice of the loop in the manifold  of the  parameters of the external driving field: by choosing the optimal loop which minimizes the ``error''  (properly estimated in terms of average fidelity loss). 
Our result is based on the  observation that there are different loops in the parameter manifold producing the same gate and, since  decoherence and dissipation crucially depend on the dynamics,    it is possible to drive the system over trajectories which are less perturbed by the 
noise.  For a simplified, albeit realistic model, we find the surprising result that the error  decreases linearly as the gating time increases.  Thus  the disturbance of the environment can be drastically  reduced. On the basis of this result, we compare holonomic  gates with the STIRAP  gates.

In Sec. II the model is introduced  and the explicit expression of the
error is derived.
In Sec. III we find the optimal loop, calculate the error, make a comparison with other 
approaches, and briefly sketch how to treat  a different coupling with the environment.


\section{Model}
\label{model}
The physical model is given by three 
degenerate (or quasidegenerate) states, 
$|+\rangle$, $|-\rangle$, and $|0\rangle$, optically connected to another state $|G\rangle$.
The system is driven by lasers with different frequencies and polarizations, 
acting selectively on the degenerate states. 
This model describes various quantum systems 
interacting with a laser radiation, 
ranging from semiconductor quantum dots, such as excitons \cite{paper1-2} and 
spin-degenerate electron states \cite{troiani-molinari},
to trapped ions \cite{HQC_proposal1} or neutral atoms \cite{HQC_proposal}.


The (approximate) Hamiltonian  modeling the  effect of the laser on the 
system is (for simplicity, $\hbar=1$) \cite{HQC_proposal1, paper1-2}
\begin{equation}
\label{eq:system_hamiltonian}
      H_0(t)= \sum_{j=+,-,0} \big[ \epsilon |j\rangle \langle j| +
      (e^{-i \epsilon t} \Omega_j(t) |j\rangle \langle G| + H.c)\big]
 \mathbf{,}
\end{equation}
where $\Omega_j(t)$ are the timedependent Rabi frequencies depending on 
controllable parameters, such as the phase and intensity of the lasers, 
and $\epsilon$ is the energy of the degenerate electron states.
The Rabi frequencies are modulated within the adiabatic time $t_{ad}$, 
(which coincides with the gating time), 
to produce a loop in the parameter space  and thereby realize the periodic 
condition $H_0(t_{ad})=H_0(0)$.


The Hamiltonian (\ref{eq:system_hamiltonian}) has four time dependent 
eigenstates:
two eigenstates $|E_{i}(t)\rangle$ , $i=1,2$, called bright states, 
and two eigenstates  $|E_i(t)\rangle$, $i=3,4$, 
called dark states. The two dark states have degenerate eigenvalue 
$\epsilon$ and the two bright states have timedependent energies 
$\lambda_\pm(t) = [\epsilon \pm \sqrt{\epsilon^2 + 4 \Omega^2(t)}]/2$ with
$\Omega^2(t)= \sum_{i=\pm,0} |\Omega_i(t)|^2$ \cite{dark-bright}.


%
%
The evolution of the state is  generated by 
\begin{equation}\label{eq:U_t}
 U_t=T e^{-i\int^t_0 dt'H_0(t')},
\end{equation} 
where $T$ is the time-ordered operator.
In the adiabatic approximation, the evolution
of the state takes place in the degenerate subspace generated by
 $|+ \rangle$, $|- \rangle$, and $|0 \rangle$.
This approximation allows to separate the dynamic contribution and 
the geometric contribution from the evolution operator.
Expanding $U_t$
in the basis of instantaneous eigenstates of $H_0(t)$ (the bright and
dark states), in the adiabatic approximation, we have
\begin{equation}
	U_t\cong \sum_j e^{-i\int^t_0 E_j(t')dt'}|E_j(t)\ra\la E_j(t)|~\mathcal{U}_t,
	\label{eq:hol_operator}
\end{equation}
where
\begin{equation}
	\mathcal{U}_t=Te^{\int_0^t d \tau V(\tau)},
	\label{eq:hol_operator}
\end{equation} 
here $V$ is the operator 
with matrix elements $V_{ij}(t) = \langle E_i(t)| \partial_t | E_j(t) \rangle$.
The unitary operator $\mathcal{U}_t$ 
plays the role of timedependent holonomic
operator and  
is the fundamental ingredient for realizing complex geometric 
transformation whereas $ \sum_j e^{-i\int^t_0 E_j(t')dt'}|E_j(t)\ra\la E_j(t)|$ 
is the dynamic contribution.

Consider $\mathcal{U}_t$ for a closed loop, i.e., for $t=t_{ad}$,
\begin{equation} \label{eq:log_operator}
\mathcal{U}=\mathcal{U}_{t_{ad}}.
\end{equation}
If the initial state $|\psi_0\ra$ is a superposition of $|+\ra$ and $|-\ra$,
then $\mathcal{U}|\psi_0\ra$ is still a superposition of the same vectors
(in general, with different coefficients)\cite{paper1-2}.
Thus the space spanned by $|+\ra$ and $|-\ra$ can be regarded as 
the ``logical space'' on which the ``logical operator'' $\mathcal{U}$ acts
as a ``quantum gate'' operator.
Note that for $t<t_{ad}$, $\mathcal{U}_t|\psi_0\ra$ has, in general, also a
component along $|0\ra$. However, as it is easy to show \cite{dark-bright},
at any instant $t<t_{ad}$, $\mathcal{U}_t|\psi_0\ra$ can be expanded in the
twodimensional space spanned by the dark states $|E_3(t)\ra$ 
and $|E_4(t)\ra$. 
It is important to observe that $\mathcal{U}$  depends
only on  global geometric features of the path in the parameter manifold and 
not on the details of the dynamical evolution \cite{HQC,paper1-2}. 


To construct a complete set of holonomic quantum gates, it is sufficient 
to restrict the Rabi frequencies $\Omega_j (t)$ in such a way that 
the norm $\Omega$ of the 
vector $\vec{\Omega}=[\Omega_0 (t),\Omega_+ (t),\Omega_- (t)]$  
is time independent and the vector lies on a real three dimensional sphere \cite{HQC_proposal1,paper1-2}.
We parametrize the evolution on this sphere 
as $\Omega_+(t)= \sin ~\theta(t) \cos ~\phi(t)$,
$\Omega_-(t)= \sin ~\theta(t) \sin ~\phi(t)$ 
and $\Omega_0(t)= \cos ~\theta(t)$ with fixed
initial (and final) point in $\theta(0)=0$, the north pole
By straightforward calculation we obtain the analytical expression 
for $V(t)$
in eq. (\ref{eq:hol_operator}), $V(t)=i \sigma_y ~cos[\theta(t)]\dot{\phi}(t)$, where $\sigma_y$ is the usual Pauli matrix 
written in the basis
of dark states.
Thus, the operator (\ref{eq:hol_operator}) becomes
$\mathcal{U}_t=\cos [a(t)] -i \sigma_y ~\sin [a(t)]$,
here $a(t) = \int_0^{t} d\tau \dot{\phi}(\tau) \cos ~\theta(\tau)$.
Accordingly, the logical operator $\mathcal{U}$ (\ref{eq:log_operator}) is
\begin{equation}
\mathcal{U}=\cos~a -i \sigma_y \sin~a,
\end{equation}
where 
\begin{equation}
  a=a(t_{ad})=\int_0^{t_{ad}} d\tau \dot{\phi}(\tau) \cos ~\theta(\tau)
  \label{eq:solid_angle}
\end{equation}
is the solid angle spanned on the sphere during the evolution.
Note that the are many paths on the sphere which generate the 
same logical operator $\mathcal{U}$, and span the same solid angle $a$.


In a previous work we have studied how interaction with the environment 
disturbs the logical operator $\mathcal{U}$ \cite{hqc_noise}.
The goal of the present paper is to analyze whether and how such a disturbance
can be minimized for a given $\mathcal{U}$.
To this end,  
we model the environment as a thermal bath of harmonic
oscillators with linear coupling between 
system and environment \cite{caldeira-leggett}. 
The total Hamiltonian is then
\begin{equation}
  \label{eq:bagno}
  H =        H_0(t) +
  \sum_{\alpha=1}^N (\frac{p^2_{\alpha}}{2 m_{\alpha}} + \frac{1}{2} 
m_{\alpha} \omega_{\alpha}^2 x_{\alpha}^2 + c_{\alpha} x_{\alpha} A),
\end{equation}
where $A$ is the system interaction operator called, from now on,
 noise operator. 


We now consider the time evolution of the reduced density matrix 
of the system, determined by the Hamiltonian (\ref{eq:bagno}). We rely on the 
standard methods of the ``master equation approach,'' with the environment  
 treated in the Born approximation
and assumed to be at each time in its own thermal 
equilibrium state at temperature $T$.  
This allows to include 
the effect of the environment in the correlation  function ($k_B=1$)
\begin{equation}
  g(\tau) = \int^\infty_0 J(\omega) \bigg[ \coth\bigg(\frac{\omega}{2 T}\bigg) 
    \cos(\omega \tau) - i~ \sin(\omega \tau)\bigg] d\omega.
  \label{eq:autocorrelation}
\end{equation}
Here the spectral density is

\begin{equation}
 J(\omega)=\frac{\pi}{2}\sum_{\alpha=1}^{N}\frac{c^2_{\alpha}}{m_{\alpha}\omega_{\alpha}}  
\delta(\omega-\omega_{\alpha}),
\end{equation}
\normalsize at the low frequencies regimes, is proportional to 
$\omega^s$, with $s\geq 0$, i.e., $s=1$ describes a Ohmic 
environments, typical of baths of conduction electrons, $s=3$ describes
a super-Ohmic environment, typical 
of baths of phonons \cite{weiss,hqc_noise}. 
The asymptotic decay  of the real part of $g(\tau)$
defines the characteristic memory time of the environment.
Denoting with $\tilde{\rho}(t)$ the time evolution of the 
reduced density matrix of the system in  the interaction picture, e.g.,
$\tilde{\rho}(t) = U_t^{\dagger} \rho U_t$,
one has \cite{weiss}

{\small
\begin{eqnarray}
  \tilde{\rho}(t_{ad})& =&\rho(0)+ \nonumber \\
   &-& i\int_0^{t_{ad}} dt\int_0^t d\tau 
    \{ g(\tau) [ 
  \tilde{A} \tilde{A}^\prime \tilde{\rho}(t-\tau) - 
  \tilde{A}^\prime \tilde{\rho}(t-\tau) \tilde{A} ] \nonumber \\
   &+& g(-\tau) [
  \tilde{\rho}(t-\tau) \tilde{A}^\prime \tilde{A} - 
  \tilde{A} \tilde{\rho}(t-\tau) \tilde{A}^\prime ].
 \label{eq:non_markov_{max}_eq}
\end{eqnarray}
}
\hspace{-0.22cm} Here $\tilde{A}$ and $\tilde{A}^\prime$ stand for 
$\tilde{A}(t)$ and $\tilde{A}(t-\tau)$, with the tilde denoting
the time evolution in the interaction picture. 


In quantum information the quality of a gate 
is usually evaluated by the fidelity $\mathcal{F}$, which measures 
the closeness
between the unperturbed state and the final state, 
\begin{equation}\label{eq:fidelity}
\mathcal{F}=\la\psi_0(0)|\mathcal{U}^{\dag}\rho (t_{ad})\mathcal{U}|\psi_0(0)\ra, 
\end{equation}
where $|\psi_0(0)\ra$ is the initial state, and 
$\rho(t_{ad})=\mathcal{U}\tilde{\rho}(t_{ad})\mathcal{U}^{\dag}$ is
the reduced density matrix in the Schr\"odinger picture
starting from the initial condition $\rho(0)=|\psi_0(0)\ra\la\psi_0(0)|$.
%
%
The average error is defined as the average fidelity loss, i.e.,
\begin{equation}\label{eq:def_delta} 
\delta=<1- \mathcal{F}> = 
1-<\la\psi_0(0)|\tilde{\rho}(t_{ad})|\psi_0(0)\ra>,  
\end{equation}
where  $<\cdots>$ denotes averaging with respect to the uniform 
distribution over the 
initial state $|\psi_0(0)\ra$.


The right-handside of Eq. (\ref{eq:def_delta}) can be computed by the following steps:

\noindent (1) solving Eq. (\ref{eq:non_markov_{max}_eq}) 
in    strictly second order 
approximation; this approximation corresponds to replace 
$\tilde{\rho}(t-\tau)$ with $\rho(0)$; 

\noindent (2) using the adiabatic approximation 
$U(t-\tau,t) \approx \exp(i \tau H_0(t))$; 

\noindent (3) expanding the scalar product in
 Eq. (\ref{eq:def_delta}) with respect to a complete orthonormal 
basis $\{|\varphi_n(t)\ra\}$, $n=1,2,3$,  orthogonal to $|\psi_0(t)\ra$.
%
In this way, one obtains

\begin{equation} 
  \delta = \Bigg< \sum_{n=1}^{3} \int_0^{t_{ad}} d t~ G(t) 
   |\langle \psi_0(t) | A |\varphi_n(t) \rangle|^2 \Bigg>, 
  \label{eq:error}
\end{equation}
where

\begin{equation}
  G(t) = \int_0^t d \tau 
   \big\{ \mbox{Re}[g(\tau)] \cos(\omega_{0n} t) 
     + \mbox{Im}[g(\tau)] \sin(\omega_{0n} t))\big\}.
  \label{eq:G_t}
\end{equation}
Here, $\omega_{0n}=\omega_0-\omega_n$ are the energy differences associated 
to the transition
$\psi_0 \leftrightarrow \phi_n$,
with $\omega_0=\epsilon$, $\omega_1=\lambda_+$,
 $\omega_2=\lambda_-$, and  $\omega_3=\epsilon$.



The interaction between system and environment is 
expressed by  the noise operator $A$ in Eq. (\ref{eq:bagno}).
We shall now make the assumption that $A=\mbox{diag}\{0,0,0,1 \}$ in the 
$| G\rangle$, $| \pm \rangle$, and $| 0 \rangle$ basis.
In this case the transition between degenerate states are forbidden, 
however the noise breaks their degeneracy, shifting one of them.
In spite of its simple form,
this $A$ is nevertheless a realistic noise operator 
for physical semiconductor systems
\cite{roszak}.


\section{Minimizing the error}

The problem can be stated in the following way: given the noise operator 
$A$ and the logical operator $\mathcal{U}$, 
find a path on the parameter space 
(the surface of the sphere, described above)
which minimizes the error $\delta$.


The total error $\delta$, given by  Eq. (\ref{eq:error}), 
can be decomposed as 
\begin{equation}
 \delta=\delta_{tr}+\delta_{pd},
\end{equation}
where the transition error, $\delta_{tr}$, is the contribution to the sum of the nondegenerate states ($\omega_{0n}\neq0$) and the pure dephasing error $\delta_{pd}$
is the contribution of the degenerate states ($\omega_{0n}=0$).
 Thus 
 
 \begin{eqnarray}\label{eq:errore_PD}
  \delta_{pd}&=&\frac{\pi}{8}\int_0^{t_{ad}}dt \int_0^{\infty} d\omega 
             \frac{J(\omega)}{\omega} \coth\bigg(\frac{\omega}{2 T}\bigg) \nonumber \\ 
           && \sin(\omega t)
              \bigg(1+\frac{1}{2}\sin^2 ~2a(t)\bigg)\sin^4~\theta(t) 
 \end{eqnarray}
and 
\begin{equation}\label{eq:errore_tr}
\delta_{tr}=\sum_{n=+,-}\frac{1}{8\sqrt{1+[(\lambda_n-\epsilon)/\Omega}]^2}\Gamma_{0n}\int_0^{t_{ad}}\sin^2 ~2\theta(t) dt ,
\end{equation}
where 
\begin{equation}
  \Gamma_{0n} = J(|\omega_{0n}|) \bigg[\coth \left( \frac{|\omega_{0n}|}{2 T} \right) - \textrm{sgn}(\omega_{0n})\bigg] 
\end{equation}
correspond to the transition rates calculated by standard Fermi golden rules, 
supposing, as usual, $G(t)\approx G(\infty)$ for $g(\tau)$ strongly 
peaked around $\tau=0$. 
In the following we define for simplicity 
\begin{displaymath}
K=\sum_{n=+,-}\frac{1}{8\sqrt{1+[(\lambda_n-\epsilon)/\Omega}]^2}\Gamma_{0n}.
\end{displaymath}

Since  we are interested at long time evolution, we start discussing 
the transition error which dominates in this regime \cite{roszak,alicki}.

\begin{figure}
    \begin{center}
         {\includegraphics[height=4.6cm]{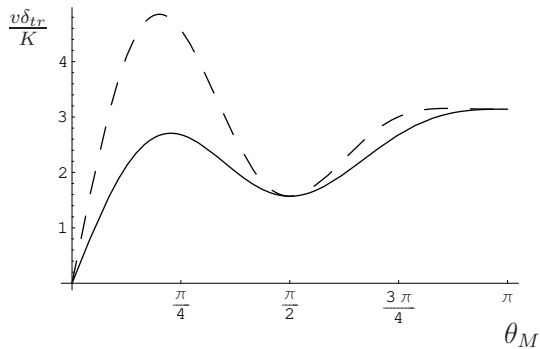}}


   \end{center}
 \caption{\label{fig:d_theta}
  The error  $\delta_{tr}$ versus $\theta_M$ for two
  different $a$ values: $a=\pi/2$ (dashed line) and $a=\pi/4$ (full line)
  correspond to NOT and Hadamard gate, respectively.
}
	
 \end{figure}


\subsection{Transition rate}
As explained in Sec. \ref{model}, the holonomic paths are 
closed curves on the surface of the sphere 
which start from the north pole.
It turns out that the curve minimizing  
$\delta_{tr}$ can be found among the loops which are composed 
by a simple sequence of three paths (see the Appendix): 
evolution along a meridian ($\phi=\mbox{const}$), evolution along 
a parallel ($\theta=\mbox{const}$) and a final evolution along 
a meridian to come back to the north pole.


The error $\delta_{tr}$ in (\ref{eq:errore_tr}), 
 depends on 
 $a$ given by Eq. (\ref{eq:solid_angle}), $\theta_M$  (the maximum angle 
spanned during the evolution along the meridian), $\Delta \phi$ (the angle 
spanned along the parallel), and angular velocity $v$. 
We allow $\Delta \phi \geq 2 \pi$ which corresponds to cover more than one 
loop along the parallel.
The velocity along the parallel is $v(t)=\dot{\phi}(t) \sin~\theta$  and 
that along the meridian is $v(t)= \dot{\theta}(t)$.
In the following we assume that $v$ is constant, and it cannot exceed 
the maximal value of $v_{\mbox{max}}$, 
fixed by adiabatic condition $v_{\mbox{max}}\ll \Omega$.


The parameters $a$, $\theta_M$, and $\Delta\phi$ 
are connected by the relation $a=\Delta \phi(1-\cos~\theta_M)$. 
The error $\delta_{tr}$ is then

\begin{equation}
  \delta_{tr}= \delta^M_{tr}+ \delta^P_{tr},
 \label{eq:f}
 \end{equation} 
where
 \begin{equation}
  \delta^M_{tr}=\frac{K}{v}\bigg(\theta_M-\frac{1}{4}\sin 4\theta_M\bigg)
  	\label{eq:fM}
\end{equation}  
is the contribution along the meridian and 
\begin{equation}
  \delta^P_{tr}=K\frac{a}{v}\frac{\sin~\theta_M~\sin^2 ~2\theta_M}{1-\cos~\theta_M}
  	\label{eq:fP}
 \end{equation}
is the contribution along the parallel.


In Fig.~\ref{fig:d_theta} $\delta_{tr}$ is plotted for 
$a=\pi/2$ and $a=\pi/4$ (corresponding to NOT and Hadamard gate, respectively) 
as a function of $\theta_M$. 
One can see that $\delta_{tr}$ has a local minimum for $\theta_M=\pi/2$ and 
a global minimum for $\theta_M=0$ where the error vanishes. 
This suggests that the best choice is to take $\theta_M$ as small as possible.


It is interesting to consider the dependence of $\delta_{tr}$ also
on the evolution time $t_{ad}$. For simplicity, we set the velocity 
$v=v_{\mbox{max}}$. In this case,
changing $\theta_M$ (and then $\Delta \phi$) corresponds to a change 
in the evolution time. We obtain 
\begin{equation}
 \theta_M=\arccos \bigg( 1-\frac{a}{2\pi m} \bigg),
\label{eq:theta-t}
\end{equation}
where
\begin{equation}
 m=\frac{1}{4\pi a}\big[(v_{\mbox{max}} t_{ad})^2+a^2\big].
\label{eq:m-t}
\end{equation}
Using these relations, $\delta^M_{tr}$ and $\delta^P_{tr}$,  given 
by (\ref{eq:fM}) and (\ref{eq:fP}) become functions of $t_{ad}$, $v_{\mbox{max}}$, and $a$.
Note that $m$ measures the space covered along the parallel, 
in fact $\Delta\phi=2\pi m$.
%

\begin{figure}
    \begin{center}
       {\includegraphics[height=4.6cm]{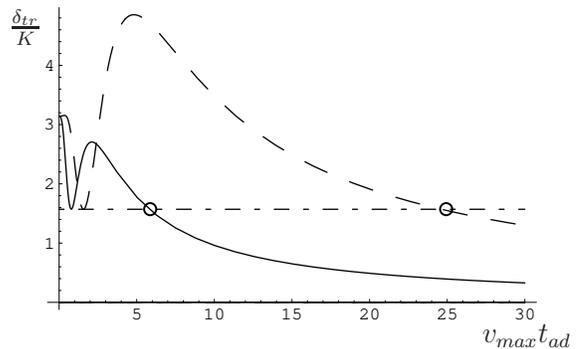}}
   \end{center}
 \caption{\label{fig:d_t} 
  The error $\delta_{tr}$ versus  $v_{\mbox{max}} t_{ad}$
  for two
  different $a$ values: $a=\pi/2$ (dashed line) and $a=\pi/4$ (full line)
  correspond to NOT and Hadamard gate, respectively.
  The dotted-dashed line shows the value of the error at $\theta=\pi/2$.
 The circles show the critical value of $v_{\mbox{max}} t_{ad}$ 
 above which the best loop is the one with  the minimal $\theta_M$.
 }
 \end{figure}

In Fig.~\ref{fig:d_t}
we see the behavior of $\delta_{tr}$ as a function of $v_{\mbox{max}} t_{ad}$.
The first minimum for both curves corresponds to $\theta_M=\pi/2$, 
then the curves for long $t_{ad}$ decrease asymptotically 
to zero corresponding
to the region in which $\theta_M \rightarrow 0$.
In this regime we have $\delta_{tr}\propto 1/t_{ad}$ which is 
drastically different from the results obtained with other methods 
where $\delta_{tr}\propto t_{ad}$,
(see Refs \cite{alicki,roszak} and below Sec. \ref{sec:comparison}). 
It should be observed that this surprising results is a merit of holonomic 
approach which allows  to choose the loop in the parameter 
space, without changing the 
logical operation as long as it subtends the same solid angle. 
Observe that small $\theta_M$ and long $t_{ad}$ mean large value of $m$, 
i.e., multiple loops around the north pole.


Figure \ref{fig:d_t} shows that, for a given gate, there is 
a critical value $k_c$ of 
$v_{\mbox{max}} t_{ad}$ which discriminate between the choice of $\theta_M$
(e.g., $k= 6$ for the Hadamard gate and $k = 25$ for the NOT gate).
For $v_{\mbox{max}}t_{ad}< k_c$ the best choice for the loop is $\theta_M = \pi/2$;
For $v_{\mbox{max}}t_{ad}> k_c$ the best choice is the  value 
of $\theta_M$ determined by 
eq. (\ref{eq:theta-t}) and (\ref{eq:m-t}).


Note that the region $v_{\mbox{max}}t_{ad}> k_c$  
is accessible with physical realistic parameters \cite{paper1-2}. For example,
if we choose the laser intensity $\Omega=20$ meV and  $v_{\mbox{max}}=\Omega/50$ 
(for which values the nonadiabatic transitions are forbidden),
the critical parameter corresponds to the critical time of $15$ ps 
for the Hadamard gate 
and $42$ ps for the NOT gate. 


\subsection{Pure Dephasing}
Until now  we have ignored the pure dephasing effect because we have 
assumed that it is 
negligible in comparison with the transition error for long evolution time. 
Now, we check that the pure dephasing error contribution can indeed be neglected. 
We can write the pure dephasing error using 
Eq. (\ref{eq:errore_PD})  
and splitting to parallel and meridian part as
%
\begin{eqnarray}\label{eq:errore_pdP}
&&\delta_{pd}^P=\int_0^{t_{ad}}dt\int_0^{\infty}d\omega\frac{J(\omega)}{\omega} 
              \coth \bigg(\frac{\omega}{2 T}\bigg) \nonumber \\
         && Q[a(t)]\sin~\omega t ~\sin^4~\theta_M
\end{eqnarray}

and

\begin{eqnarray}\label{eq:errore_pdL}
&&\delta_{pd}^M=\int_{0}^{\frac{\theta_M}{v_{\mbox{max}}}}dt
\int_0^{\infty}d\omega\frac{J(\omega)}{\omega}\coth \bigg(\frac{\omega}{2 T}\bigg) Q[a(t)]\nonumber \\
&& \sin~\omega t\bigg\{\sin^4(v_{\mbox{max}} t)+\sin^4\bigg[\theta_M\bigg(1-\frac{v_{max} t}{\theta_M}\bigg)\bigg]\bigg\},
\end{eqnarray}
where $Q[a(t)]= 1+ 1/2 ~\sin^2[2 a(t)]$.

To estimate $\delta_{pd}$ we assume that $t_{ad}$ is longer with respect to 
the characteristic time of the bath. 
Remembering that $J(\omega)\propto \omega^s$, 
the pure dephasing error behavior along the parallel part at the temperature $T$ 
is

\begin{equation}
\delta^P_{pd}\propto\left\{\begin{array}{lr}
 \left(\frac{\displaystyle 1}{\displaystyle t_{ad}}\right)^{s+3}, & T\ll1/t_{ad} \\
 T \left(\frac{\displaystyle 1}{\displaystyle t_{ad}}\right)^{s+2}, & T\gg1/t_{ad}
\end{array}\right.
\end{equation}  
 while the along meridian is
\begin{equation}
 \delta_{pd}^M \propto \left(\frac{1}{t_{ad}}\right)^3. 
\end{equation}
 Then, we can conclude that the pure dephasing 
can always be neglected for long time evolution because
 it decreases faster than the transition error.

\subsection{Comparison between HQC and STIRAP}
\label{sec:comparison}
We make a comparison between holonomic quantum computation (HQC) 
and the STIRAP procedure
which is an analogous approach to process quantum information.
The STIRAP procedure (\cite{Kis, roszak}) is, in its basic points, 
very similar to 
the holonomic information manipulation. 
The level spectrum, the information encoding,
the evolution produced by adiabatic evolving laser are exactly the same.
The fundamental difference is that in STIRAP 
the dynamical evolution is fixed
(we must pass through a precise sequence of states) 
and then the corresponding loop in the parameter space is fixed.
In particular, we go from the north pole to the south pole 
and back to the north pole
along meridians. Since the loop, as in our model, is a 
sequence of meridian-parallel-meridian path,
we can calculate the error and make a direct comparison.
In this case, the transition error 
results proportional to $\delta_{tr}\propto t_{ad}$
and grows linearly in time while for HQC $\delta_{tr}\propto1/t_{ad}$.
Therefore, the HQC is fundamentally the favorite for 
long application times with respect to the STIRAP ones.

Moreover, we can show that the freedom in the choice of the loop 
allows us to construct HQC 
which perform better than the best STIRAP gates.
In Ref. \cite{roszak} the minimum error (not depending on the evolution time) 
for STIRAP was obtained reaching a compromise between the necessity 
to minimize the transition, pure dephasing error and the constraint 
of adiabatic evolution.
With realistic physical parameters \cite{hqc_noise} ($J(\omega)=k\omega^3 e^{(-\omega/\omega_c)^2}$, $\Omega=10$ meV, $\epsilon=1$eV, $v_{max}=\Omega/50$, $k=10^{-2} ($meV$)^{-2}, \omega_c=0.5$ meV and for low temperature), 
the total minimum error 
in Ref. \cite{roszak} is $\delta_{\mbox{stirap}}= 10^{-3}$.
With the same parameters, we still have the possibility to increase the evolution time in order to 
reduce the environmental error. However, for evolution time $t_{ad}=50$ ps we obtain a total error
$\delta= 1.5 \times 10^{-4}$ for the NOT gate and $\delta = 4 \times 10^{-5}$ for the Hadamard gate, respectively.
As can be seen, the logical gate performance is greatly increased.

\subsection{More general noise}
 Until now we have discussed the possibility to minimize the environmental error by
 choosing a particular loop in the parameter sphere but the structure of the error functional clearly depends on the system-environment interaction.
 Then one might wonder if the same approach can be used for a different noise environment.

 For this reason, we now briefly analyze the case of noise matrix in the form 
 $A=\mbox{diag}\{0,1,0,-1 \}$. 
 Again, for long evolution we can neglect the contribution of the 
pure dephasing and focus on the transition 
 error. In this case the interesting part of the error 
functional takes the form 

 \begin{equation}
  \delta_{tr}=K[(\frac{1}{2}\sin ~2\theta ~\cos ~2\theta)^2+(\sin~\theta~\sin ~2\phi)^2].
 \end{equation}

Even if the analysis in this case is much more complicated, 
it can be seen that $\delta_{tr}$ has an 
absolute minimum for $\theta_M=0$.
The long time behavior is the same 
($\delta_{tr} \propto 1/ t_{ad}$) such that the results 
are qualitatively analogous to the above ones:
for small $\theta_M$ loops (or long evolution at fixed velocity) the holonomic 
quantum gate presents 
a decreasing error.
Then even in this case it is possible to minimize the environmental error.

\section{Conclusions}
In summary, we have analyzed the performance of holonomic quantum gates 
in the presence of environmental noise 
by focusing on the possibility to have small errors choosing 
different loops in the parameter manifold.
Due to the geometric dependence, we can implement the same logical 
gate with different loops.
Since  different loops correspond to different dynamical evolutions, 
we have used this freedom to 
construct an evolution through ``protected'' or ``weakly  influenced'' 
states leading to good holonomic quantum gates performances.  
This allows to select (once that the physical parameter are fixed)
the best loop which minimizes the environmental effect. 
(Note that this optimization procedure is rather independent of the details
 of the simple model we have considered and arguably, it could be
extended to more complicated systems without any substantial modification.)
We have shown that for long time evolutions the noise decreases 
as $1/t_{ad}$ while in the other cases 
it increases linearly with adiabatic time.
We also have shown that the same features can be found with 
different kinds of noise suggesting 
the possibility to find a way to minimize the environmental effect 
in the presence of any noise.
These results open a new possibility for implementation of holonomic 
quantum gates to 
build quantum computation 
because they seem robust against both control error and environmental noise.


\section*{Acknowledgment}
The autors thank E. De Vito for useful discussions.
One of the authors (P. S.) acknowledges support from INFN.
Financial support by the italian MIUR via PRIN05 and INFN is acknowledged.

\appendix
\section{Minimizing theorem}\label{app:proof}

Let us consider the family $\mathcal{C}_n$ composed of the closed curves 
generated by a sequence of $n$ paths along a  parallel 
($\theta=const$) alternated with paths along a  meridian ($\phi=const$).
We call  $C_n$ a generic curve in this family. 
For example, the family $\mathcal{C}_1$ contains all the closed 
curves composed by the sequence
of path  meridian-parallel-meridian while the 
family $\mathcal{C}_2$ 
contains the curves 
meridian-parallel-meridian-parallel-meridian.

We argue that the closed curve minimizing the 
error in Eq. (\ref{eq:errore_tr}) can be found in the 
$\mathcal{C}_1$ family.
First, we show that any closed curve in $\mathcal{C}_2$ spanning 
a solid angle $a$ on the sphere
can be replaced by a closed curve in $\mathcal{C}_1$ spanning 
the same angle and producing a smaller error.
In analogous way any closed curve in $\mathcal{C}_3$ 
can be replaced by a closed curve in 
$\mathcal{C}_2$ with smaller error  and so on.
By induction we obtain that any closed curve in $\mathcal{C}_n$ 
can be replaced by a curve in 
$\mathcal{C}_1$ spanning the same solid angle but producing smaller error.
Since the curve belonging to $\mathcal{C}_n$ can approximate 
any closed curve on the sphere,
the best curve can be found in $\mathcal{C}_1$.

The crucial point is to show that any curve in $\mathcal{C}_2$ 
can be replaced by a curve in 
$\mathcal{C}_1$. Let us consider a generic curve $C_2$ in $\mathcal{C}_2$ 
spanning a solid angle $a$: composed by a segment of a meridian 
(with $\theta$ going from $0$ to $\theta_1$),  
a parallel (spanning a $\Delta \phi_1$ angle),  meridian 
(with $\theta:\theta_1 \rightarrow \theta_2$), a  parallel
(spanning a $\Delta \phi_2$ angle), and finally a segment to the 
north pole along a meridian.
Let us consider two closed curves $C_1^1$ and $C_1^2$ in $\mathcal{C}_1$
subtending the same solid angle $a$ 
with, respectively, $\theta_1$ and $\theta_2$ as maximum angle 
spanned during the evolution along the  meridian.
First we analyze (\ref{eq:f}) along the meridian.
Without losing generality, we can take $\theta_1 < \theta_2$; 
it is clear from Eq. (\ref{eq:fM})
that the value of $\delta_{tr}$ along the meridian for $C_1^1$ 
is smaller that for $C_1^2$: 
$\delta^M_{C_1^1} < \delta^M_{C_1^2}$.
We note from the Eq. (\ref{eq:fM}), suitable extended to $C_2$,
that the two paths along the meridians
depends only on  $\theta_2$ and then produce the same error of $C_1^2$,

\begin{equation}
 \delta^M_{C_1^1} < \delta^M_{C_1^2} = \delta^M_{C_2}.
	\label{app_eq:meridian_error}
\end{equation}


The difference between the contribution along the parallel is

 \begin{equation}\label{eq:dif1}
 \delta^P_{C_2}-\delta^{P}_{C_1^2}=\Delta\phi_1\bigg(\sin~\theta_1~\sin^2~2\theta_1-\frac{1-\cos~\theta_1}{1-\cos~\theta_2}\sin~\theta_2~\sin^2~2\theta_2\bigg) 
\end{equation}
and

 \begin{equation}\label{eq:dif2}
 \delta^P_{C_2}-\delta^{P}_{C_1^1}=\Delta\phi_2 \bigg(\sin~\theta_2~\sin^2~2\theta_2-\frac{1-\cos~\theta_2}{1-\cos~\theta_1}\sin~\theta_1~\sin^2~2\theta_1\bigg).
 \end{equation}

Analysis of the positivity of the quantities 
given by Eqs. (\ref{eq:dif1}) and (\ref{eq:dif2}) shows that
 $\delta^P_{C_2}$ cannot be at the same time smaller 
than $\delta^{P}_{C_1^2}$ and $\delta^{P}_{C_2^2}$.
In fact, there are two possibilities:
If $\delta^P_{C_2} > \delta^P_{C_1^1}$, from Eq. (\ref{app_eq:meridian_error}) and (\ref{eq:dif2}),
\begin{equation}
  \delta_{C_2}= \delta^M_{C_2}+\delta^P_{C_2} > \delta^M_{C_1^1}+\delta^P_{C_1^1} = \delta_{C_1^1},
\end{equation}
and the best  closed curve is $C_1^1$.
If $\delta^P_{C_2} > \delta^P_{C_1^2}$, from Eqs. (\ref{app_eq:meridian_error}) and (\ref{eq:dif1}),
\begin{equation}
  \delta_{C_2}= \delta^M_{C_2}+\delta^P_{C_2} > \delta^M_{C_1^2}+\delta^P_{C_1^2} = \delta_{C_1^2},
\end{equation}
and the best  closed curve is $C_1^2$.

In the same way it can be shown that any closed curve in $\mathcal{C}_3$ 
can be replaced by a closed curve in $\mathcal{C}_2$ with smaller error.




\begin{thebibliography}{99}
\bibitem{dark-bright}
The explicit expression for the  bright states is
$|E_{1} \rangle = \frac{1}{\sqrt{2} \Omega}
(\Omega |e\rangle + \sum_i \Omega_i |i \rangle)$ and
$|E_{2} \rangle = \frac{1}{\sqrt{2} \Omega}
(-\Omega |e\rangle + \sum_i \Omega_i |i \rangle)$;
for the dark states is
$|E_3\rangle = 1/(\Omega \sqrt{|\Omega_+|^2+ |\Omega_-|^2})
[ \Omega_0 (\Omega_+ |+ \rangle + \Omega_- |- \rangle) -
(\Omega^2 - |\Omega_0|^2) |0 \rangle ]) $ and
 $|E_4 \rangle = 1 / \sqrt{|\Omega_+|^2 + |\Omega_-|^2}
[\Omega_- |+ \rangle - \Omega_+ |- \rangle]$.
\end{thebibliography}
\end{document}